# PulmoBell: Home-based Pulmonary Rehabilitation Assistive Technology for People with COPD

Yuanxiang Ma, Andreas Polydorides, Jitesh Joshi, and Youngjun Cho*

University College London

Abstract: Chronic Obstructive Pulmonary Disease (COPD) can be fatal and is challenging to live with due to its severe symptoms. Pulmonary rehabilitation (PR) is one of the managements means to maintain COPD in a stable status. However, implementation of PR in the UK has been challenging due to the environmental and personal barriers faced by patients, which hinder their uptake, adherence, and completion of the programmes. Moreover, increased exercise capacity following PR does not always translate into physical activity (PA) and unfortunately, can lead back to exercise capacity seen prior to PR. Current alternative solutions using telerehabilitation methods have limitations on addressing these accessibility problems, and no clear conclusion can be drawn on the efficacy of telerehabilitation in enhancing the sustainability of PR outcomes via promoting PA in patients' everyday life. In this work, the authors propose a novel design of sensor-based assistive product with the aim of facilitating PR and promoting PA maintenance in a home-based setting. Prototypes of different levels of fidelity are presented, followed by an evaluation plan for future research directions.

## 1 INTRODUCTION

Chronic Obstructive Pulmonary Disease (COPD) is a progressive condition of the respiratory system characterised by reduced air supply to the lungs [2]. The airflow limitation is directly associated with an inflammatory response in the lungs and primarily caused by long-term exposure to and consequent inhalation of noxious particles or gases [1]. COPD is responsible for over 3.23 million deaths in 2019, becoming the third leading mortality factor in the world and the second leading factor of Disability-Adjusted Life-Years (DALYs) reduction [3, 5]. An estimated 1.2 million people are diagnosed and living with COPD in the UK, accounting for around 2% of the total population [4]. Patients' exercise capacity, life quality, emotional functions, and social performance are impaired [1, 6-8]. COPD can be categorised as a hidden disability, leading to facing a wide range of barriers that may hinder people's participation in society and interaction [9].

COPD is preventable and treatable through a set of interventions such as tobacco cessation courses, pharmacological therapies, long-term oxygen therapies (LTOT), and pulmonary rehabilitation (PR) programmes [1, 11, 12]. As a significant component of these managements, PR is a comprehensive and multidisciplinary intervention of care, comprising individually-tailored exercise training, behaviour management, nutrition therapy, etc. [7, 13, 29]. While there is evidence [2] suggesting that PR has a positive effect on health-related quality of life, it is underused, with approximately 3% - 16% of COPD patients eligible for referral to the programme, and only 1-2% gaining access [7]. Additionally, around half the patients with severe and very severe symptoms show unwillingness to uptake PR programmes based in hospitals or clinics, and approximately 30-50% of those who have attended PR programmes quit before completion [8]. Thus, there is a need to facilitate the increased uptake, adherence, and completion of PR programmes by patients.

Environmental factors such as long-distance travel, non-accessible public transport, parking difficulties, inconvenient timing, seasonal weather patterns, and air pollution were considered major external causes of the non-attendance and non-completion of PR programmes in patients [7,16]. Personal and health-related problems were also listed as the primary factors affecting PR access [15,17,18,20]. Personal issues describe patients' objective living status and subjective attitudes toward PR, including worsening emotional status, unhealthy lifestyles, living alone, low expectations, negative attitudes, anxiety, and concerns. Health-related problems include fluctuations or exacerbations of COPD symptoms, such as worsening dyspnea, changing exercise capacities, developing comorbidities or new medical conditions, etc. In their review, Thorpe et al. [7] suggest that external obstacles risk aggravating personal symptoms (e.g., dyspnea, sputum production, muscle weakness), whilst the symptoms, in turn, evoke anxiety and fear, leaving patients more vulnerable to environmental factors.

* Supervisor

The literature provides evidence for understanding the treatment effects and accessibility issues of hospital-based PR programmes. However, it does not address factors influencing patients' maintenance of long-term behaviour change and physical activity (PA) following the completion of PR programmes. Physical activity following a PR programme is vital in reinforcing a stable state of COPD, and a worse prognosis is often associated with physical inactivity [26]. Although evidence [2] reveals that PR enhances exercise capacity through exercise training, the treatment outcome does not necessarily translate into increased PA and may return to its previous status prior to PR [21]. Furthermore, Robinson et al. [19] suggest that symptom-evoked anxiety, limited access to social support, and lack of positive feedback on health status are key obstacles that hinder patients from establishing routines incorporating PA and reinforcing behaviour change in everyday life. Thus, current PR may not be capable of maintaining sustainability in its outcomes, in part due to its outpatient format and multidisciplinary nature [6] within the National Health Service (NHS), which can be less flexible, takes many medical resources, and results in a lack of continuous support for the self-management of COPD post-programme.

## 2 PULMOBELL

### 2.1 Overview of the functionality

We have designed Pulmobell to address the multifaceted needs of COPD patients. Inside the housing, the weight elements are accompanied by an accelerometer, an oximeter, and an air quality sensor (PM2.5/PM10). When the air quality is good enough for COPD patients to exercise, Pulmobell can prompt users through a notification and when the air is too polluted they can be warned to avoid it or find a cleaner indoor area. The accelerometer provides data on the number of repetitions carried out by the user and can help them match their assigned routine. The respiratory rate (a key respiration metric [10]), inferred from the oximeter, along with the oxygen saturation level can be used to adjust the intensity of the exercise regimen throughout the session. The Arduino module housed within Pulmobell sends the data through Bluetooth to the mobile app so that the user can view and interact. This data can also be available to be viewed by the user's personal doctor or therapist. An overview of the system within Pulmobell is shown in Figure 1.

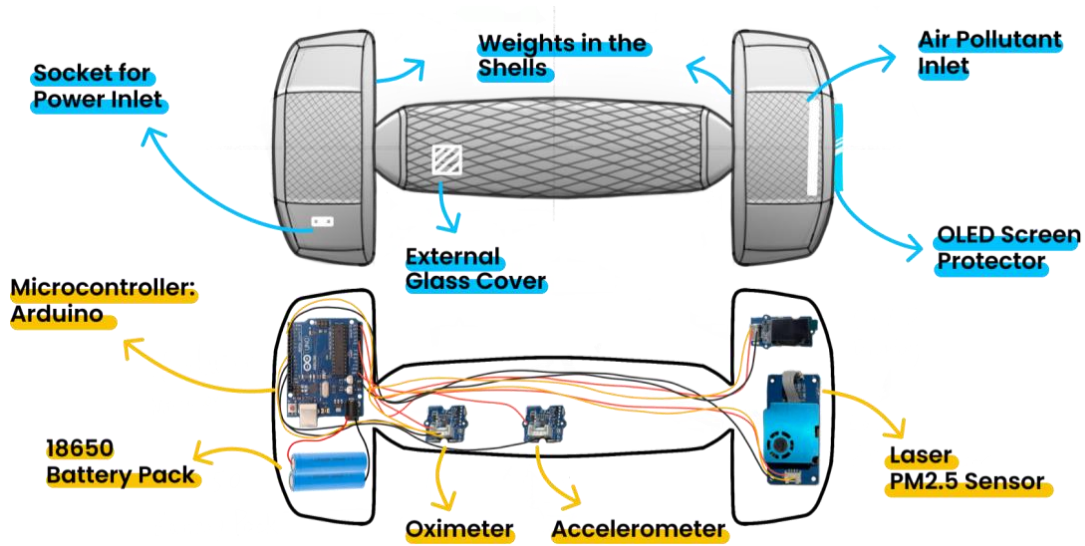

Figure 1: Front view and cross-section view of sketches of a sensor-based dumbbell for home-based PR.

### 2.2 Design Considerations

The design decisions on Pulmobell suit the requirements and needs gathered through the literature review. The variety of sensors employed are utilised to make real-time feedback recognisable to users to boost their self-efficacy on PA maintenance [19]. While a central IoT hub with sensors and displays was considered, this would be impractical as this hub would need not only be in the same room as the user but near them as they are moving around dust particles from their clothes and weights. Apart from the PM2.5 sensor, incorporating the oximeter into Pulmobell's handle allows for real-time readings and does require pausing for measurements during the session. This is important, COPD is more prevalent in an elderly population, so a streamlined exercise experience was necessary for a population that may not be as tech-savvy. Finally, resistance training was chosen as it is the mainstay of exercise training in PR, defined as *"an exercise modality in which local muscle groups are trained by repetitive lifting of relatively heavy loads"* [29].

## 2.3    Pulmobell Implementation

With our 3D CAD design and 3D printing, the components easily thread together for assembly and are to house the electronics and weight elements. The housing is made of ABS, providing significant impact resistance, a property we prioritised in the development of the housing. The handle has a surface that provides substantial grip, while the oximeter cut-out is tactile and allows for quick and correct grip where extra haptic actuators can be embedded for feedback [22, 23]. Figure 2 shows the product used. The display in figure 2 also displays the data available in the app. Figure 3 shows the final rendered Pulmobell interface.

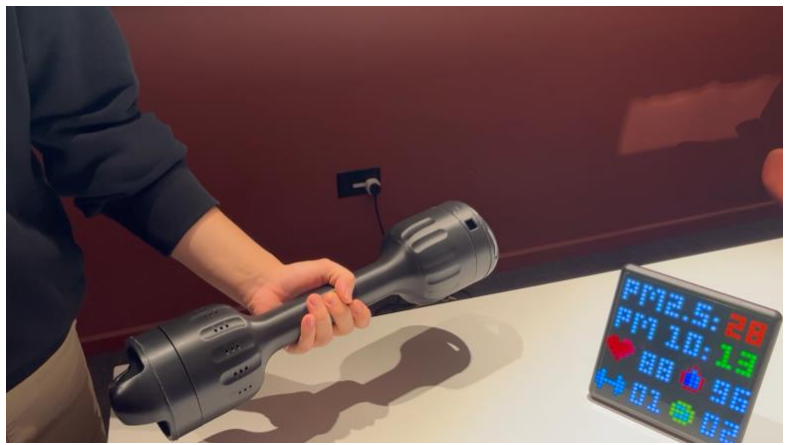

Figure 2: A user holding the Pulmobell and a display shows both health data collected and exercise regimen statistics.

## 3   DISCUSSION AND FUTURE WORK

To help address personal and environmental barriers faced by COPD patients during physical activity, this work proposes a novel interface for assisting pulmonary rehabilitation in a home-based setting, with the aim to empower the users to take up, adhere to, and complete effective exercise. In our future research, we will integrate this into physiological computing paradigms to understand how this type of assistive interface can be personalised based on the users' physiological states such as respiration [10, 24], cardiovascular states [25] (e.g., with open-source toolkits and sensors [27, 28] or thermal imaging [14, 30]) for more effective pulmonary rehabilitation programmes [31, 32]. Further, we plan to conduct interview studies with experts in COPD to develop a better understanding of how this interface can be incorporated into the mainstream pulmonary exercise practice.

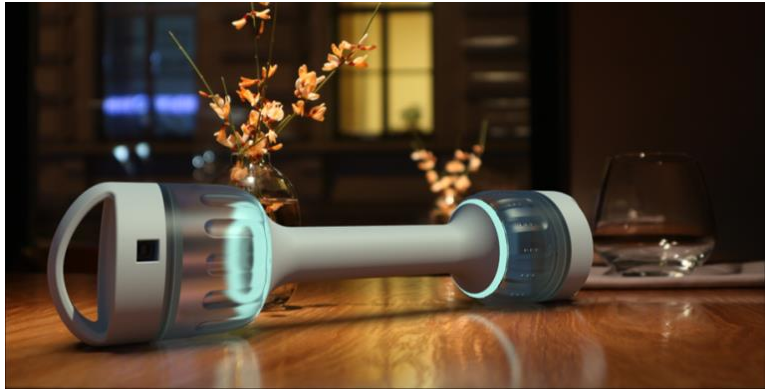

Figure 3: A rendered image of a polished exterior for Pulmobell.